\documentclass[twocolumn,twoside,slac_two]{revtex4}
\usepackage{graphicx}
\usepackage{fancyhdr}
\newcommand{\pt}{\ensuremath{p_{T}}}
\newcommand{\st}{\ensuremath{S_{T}}}

\newcommand{\pbarn} {\ensuremath{\mathrm{pb^{-1}}}}

\begin{document}

\title{ATLAS Sensitivity to Leptoquarks, $W_{\bf R}$ and Heavy Majorana Neutrinos in Final States with High-${\bf \pt}$ Dileptons and Jets with Early LHC Data at 14 TeV proton-proton collisions}

%

\author{Vikas Bansal (ATLAS Collaboration)}
\affiliation{University of Pittsburgh, 3941 O'Hara St., Pittsburgh, PA~15260, USA}

\begin{abstract}
Dilepton-jet final states are used to study physical
phenomena not predicted by the standard model. The ATLAS discovery
potential for leptoquarks and Majorana Neutrinos is presented using a
full simulation of the ATLAS detector at the Large Hadron Collider. 
The study is motivated by the role of the leptoquark in the Grand
Unification of fundamental forces and the see-saw mechanism that
could explain the masses of the observed neutrinos. The analysis algorithms
are presented, background sources are discussed and estimates of
sensitivity and the discovery potential for these processes are reported.
\end{abstract}

\maketitle

\thispagestyle{fancy}
\section{Introduction}
The Large Hadron Collider (LHC) will soon open up a new energy scale 
that will directly probe for physical phenomena outside the framework of the Standard Model (SM). 
Many SM extensions inspired by Grand Unification introduce new, very heavy particles such as leptoquarks. 
Extending the SM to a larger gauge group that includes, 
{\it e.g.} Left-Right Symmetry (LRS) \cite{Mohapatra:1975it}, 
could also explain neutrino masses via the see-saw mechanism.  
The LRS-based Left-Right Symmetric Model (LRSM) \cite{K.-Huitu:1997ye} 
used as a guide for presented studies, 
extends the electroweak gauge group of the SM 
from SU(2)$_L$ $\times$ U(1)$_Y$ to SU(2)$_L$ $\times$ SU(2)$_R$ $\times$ U(1)$_{B-L}$ 
and thereby introduces $Z^\prime$ and right-handed $W$ bosons. 
If the LRS breaking in nature is such that all neutrinos become Majoranas, 
the LRSM predicts the see-saw mechanism \cite{Mohapatra:1981oz} 
that elegantly explains the masses of the three light neutrinos. 

\section{Search for scalar leptoquarks}

\begin{figure}[htbp]
\center{
{\includegraphics[width=0.4\textwidth]{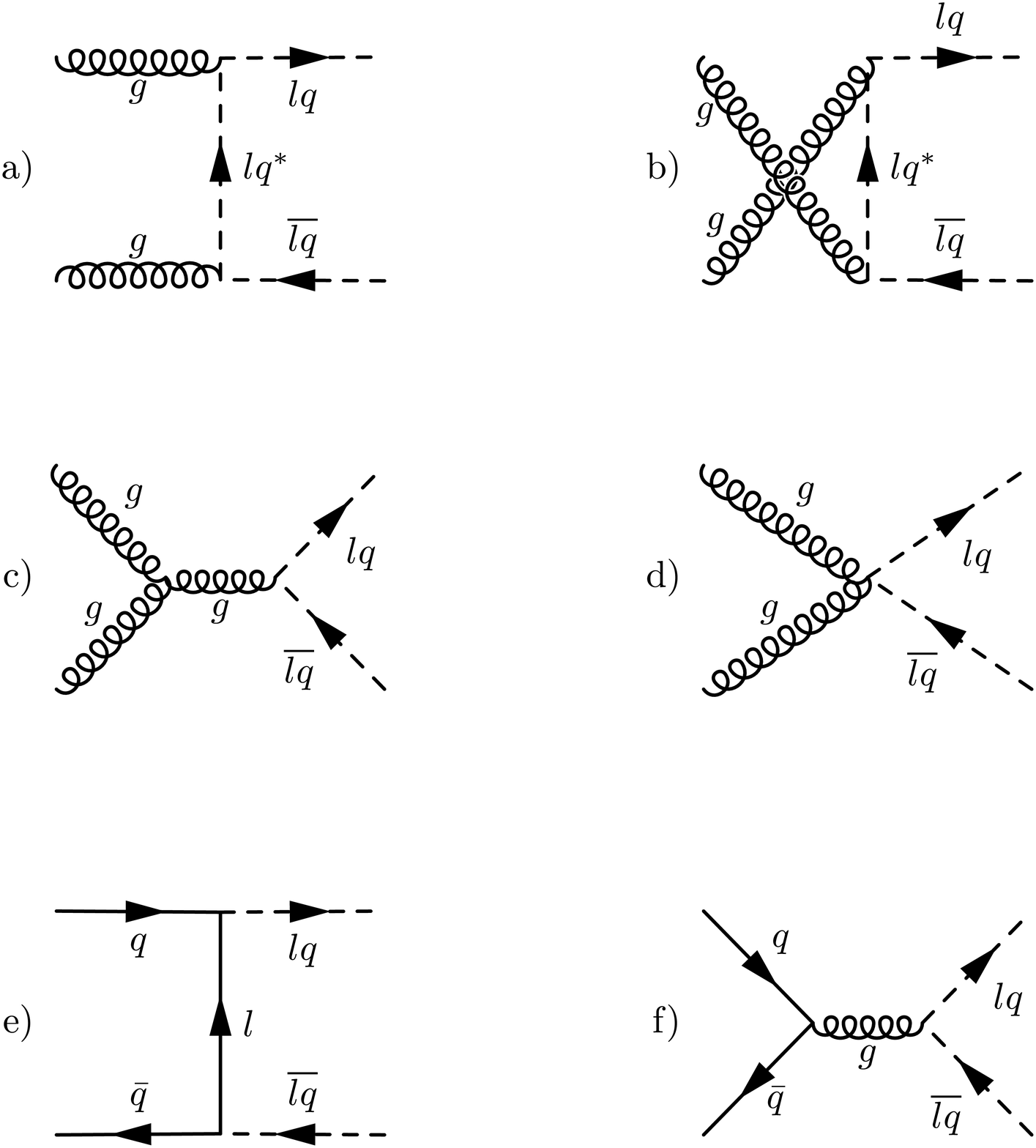}}   
}
\caption{\small {
Feynman diagrams for the pair production of scalar leptoquarks via 
	gluon-gluon fusion (a-d) and quark anti-quark annihilation (e-f). 
}}
\label{fig:LQ_feynman}
\end{figure}

Leptoquarks (LQ) are hypothetical bosons carrying both quark and lepton quantum numbers, 
as well as fractional electric charge~\cite{Buchmuller:1986iq,Georgi:1974sy}.  
Leptoquarks could, in principle, decay into any combination of any flavor lepton and any flavor quark. 
Experimental limits on lepton number violation, 
flavor-changing neutral currents, and proton decay 
favor three generations of leptoquarks. 
In this scenario, each leptoquark couples to a lepton 
and a quark from the same SM generation\cite{Leurer:1993em}. 
Leptoquarks can either be produced in pairs by the strong interaction or
in association with a lepton via the leptoquark-quark-lepton coupling. 
Figure~\ref{fig:LQ_feynman} shows Feynman diagrams for the pair production of leptoquarks at the LHC.

This contribution describes the search strategy for 
leptoquarks decaying to either an electron and a quark or a muon and 
a quark leading to final states with two leptons and at least two jets. 
The branching fraction of a leptoquark to a charged lepton and a quark is 
denoted as $\beta$\footnote{$\beta = 1$ would mean that leptoquarks do not 
decay into quarks and neutrinos.}.

MC-simulated signal events have been studied\cite{ATLAS_CSC_PHYSICS_BOOK:2008} using Monte Carlo (MC) samples for first generation (1st~gen.) 
and second generation (2nd~gen.) scalar leptoquarks simulated 
at four masses of 300~GeV, 400~GeV, 600~GeV, and 800~GeV 
with the MC generator {\sc Pythia} \cite{pythia} at 14~TeV $pp$ center-of-mass energy.
The next to leading order (NLO) cross section~\cite{Kramer:2004df} for the above simulated signal 
decreases with leptoquark mass from a few pb to a few fb with mass point of 400~GeV at (2.24$\pm$0.38)~pb.

\begin{table*}[t]
\begin{center}
\caption{Partial cross-section ({\rm pb}) that remains after each selection criterion for the first generation leptoquark channel. Baseline selection corresponds to the first selection of two electron candidates and two jets from the reconstructed objects.
The $Z$/DY cross section is for the region $M(ee) \ge 60$~GeV. 
VB pairs corresponds to the diboson processes of WW, WZ, and, ZZ.
}
\begin{tabular}{|l|c|c|c|c|c|c|}
\hline
Physics               & Before             & Baseline  & $S_T$         & $M_{ee}$       & M$_{lj}^{1}$ - M$_{lj}^{2}$ mass window\\
sample                & selection          & selection & $\ge 490$~GeV & $\ge 120$~GeV  &  [ 320-480 ] -  [ 320-480 ]  \small{[GeV]}  \\
\hline		                           
LQ (400 GeV)          & 2.24               & 1.12      & 1.07          & 1.00           &  0.534                                \\
\hline 
$Z$/DY~$\ge$~60~GeV   & 1808.              & 49.77     & 0.722         & 0.0664         & 0.0036                              \\
$t\bar{t}$            & 450.               & 3.23      & 0.298         & 0.215          & 0.0144                                \\
VB pairs              & 60.94              & 0.583     & 0.0154        & 0.0036         & 0.00048                                 \\
\hline
Multijet              & $10^{8}$           &  20.51    &  0.229        &  0.184         & 0.0                                    \\
\hline
\end{tabular}
\label{cut_flow_LQ_ee}
\end{center}
\end{table*}

\begin{table*}[t]
\begin{center}
\caption{Partial cross-section ({\rm pb}) that remains after each selection criterion for the second generation leptoquark channel. Baseline selection corresponds to the first selection of two muon candidates and two jets from the reconstructed objects.
The $Z$/DY cross section is for the region $M(\mu \mu) \ge 60$~GeV. 
VB pairs corresponds to the diboson processes of $WW$, $WZ$, and $ZZ$.
}
\begin{tabular}{|l|c|c|c|c|c|c|c|}
\hline
Physics     & Before    & Baseline  & p$^{\mu}_{T}$$\ge$60 GeV    & S$_{T}$        & $M(\mu\mu)$    & M$_{lj}$ mass window \\
sample             & selection & selection & p$^{jet}_{T}$$\ge$25 GeV    & $\ge600$~GeV  & $\ge110$~GeV  & [ 300 - 500 ]  \small{[GeV]}\\
\hline
LQ (400 GeV)       & 2.24      & 1.70      & 1.53       & 1.27      & 1.23       & 0.974        \\
\hline 
$Z$/DY$\ge$60 GeV  & 1808.     & 79.99     & 2.975      & 0.338     & 0.0611     & 0.021       \\
$t\bar{t}$         & 450.      & 4.17      & 0.698      & 0.0791    & 0.0758     & 0.0271    \\
VB pairs           & 60.94     & 0.824     & 0.0628     & 0.00846   & 0.00308    & 0.00205  \\
\hline
Multijet           & $10^{8}$  &  0.0      &  0.0       &  0.0      & 0.0        & 0.0         \\
\hline
\end{tabular}
\label{cut_flow_LQ_mm}
\end{center}
\end{table*}

\subsection{Reconstruction and objects selection}
\label{baseline_selection}

Signal reconstruction requires selection of two high quality leptons and at least two jets. Each signal jet and lepton candidate is required to have transverse momentum ($\pt)>$~20~GeV. This helps to suppress low $\pt$ background predicted by the SM. Leptons are required to have pseudorapidity $|\eta|$ below 2.5, which is the inner detector's acceptance, whereas jets are restricted to $|\eta| < 4.5$ to suppress backgrounds from underlying event and minimum bias events that dominate in the forward region of the detector. In addition, leptons are required to pass identification criteria, which, in case of electrons, are based on electromagnetic-shower shape variables in the calorimeter and, in the case of muons, are based on finding a common track in the muon spectrometer and the inner detector together with a muon isolation\footnote{$E_T^{iso}/p_T^\mu \le 0.3$, where 
         $p_T$ is muon candidate's transverse momentum and $E_T^{iso}$ is energy detected in the calorimeters 
         in a cone of $\Delta$R=$\sqrt{(\Delta\eta^2 + \Delta\phi^2)}$=0.2 around muon candidate's reconstructed trajectory.} requirement in the calorimeter. Electron candidates are also required to have a matching track in the inner detector. Furthermore, it is required that signal jet candidates are spatially separated from energy clusters in the electromagnetic calorimeter that satisfy electron identification criteria. Finally, a pair of leptoquark candidates are reconstructed from lepton-jet combinations. Given the fact that these four objects can be combined to give two pairs, the pair that has minimum mass difference between the two leptoquark candidates is assumed to be the signal.

\subsection{Background Studies}
\label{LQ_Bckg}

The main backgrounds to the signal come from $t\bar{t}$, and $Z /DY$+jets production processes. Multijet production where two jets are misidentified as electrons, represents another background to the dielectron(1st~gen.)~channel. 
In addition, minor contributions arise from diboson production. 
Other potential background sources, such as single-top production, were also studied and found to be insignificant.  

\begin{figure}[h]
\center{
{\includegraphics[width=2.9in]{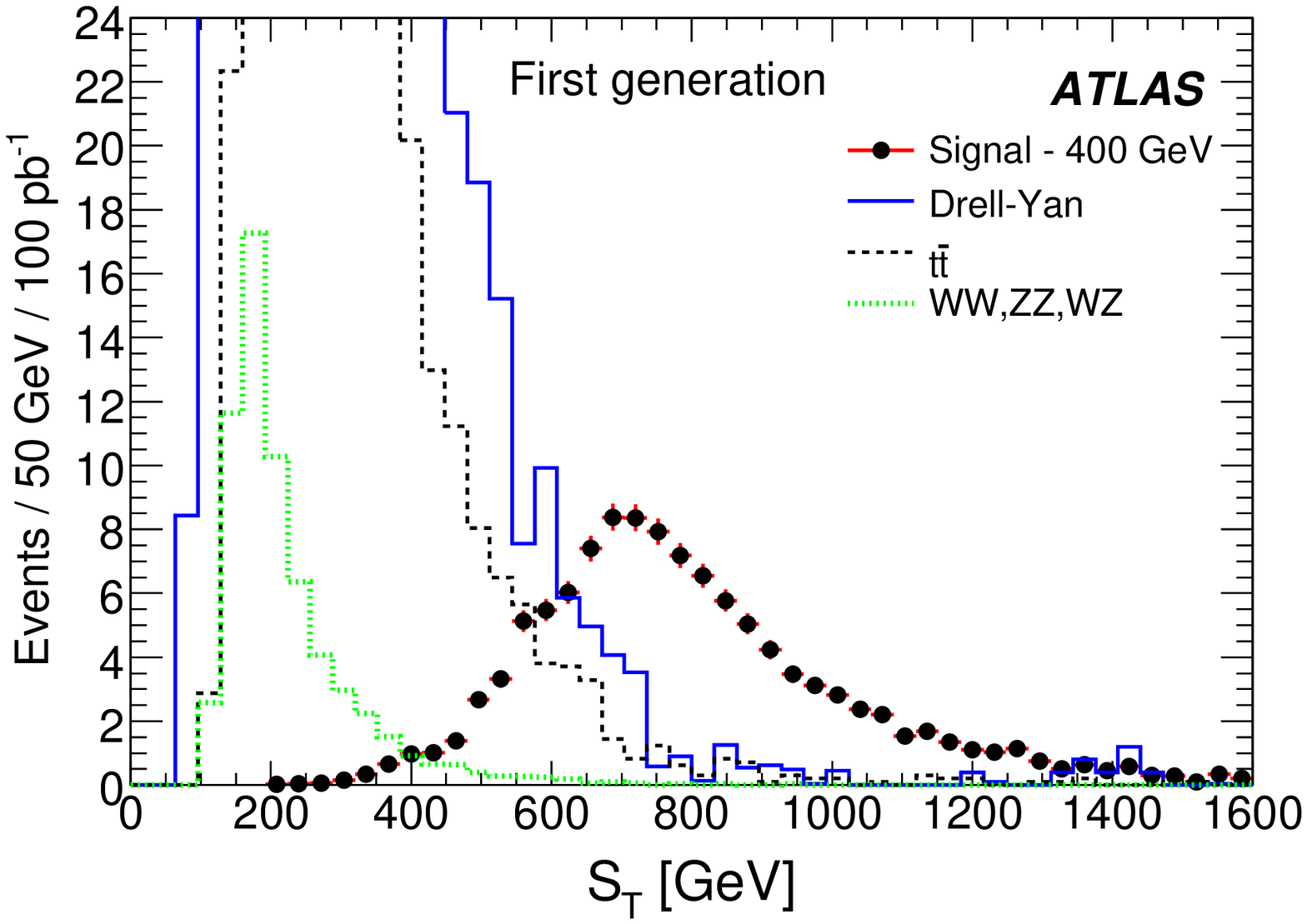}}
\hspace{0.5in}
{\includegraphics[width=2.9in]{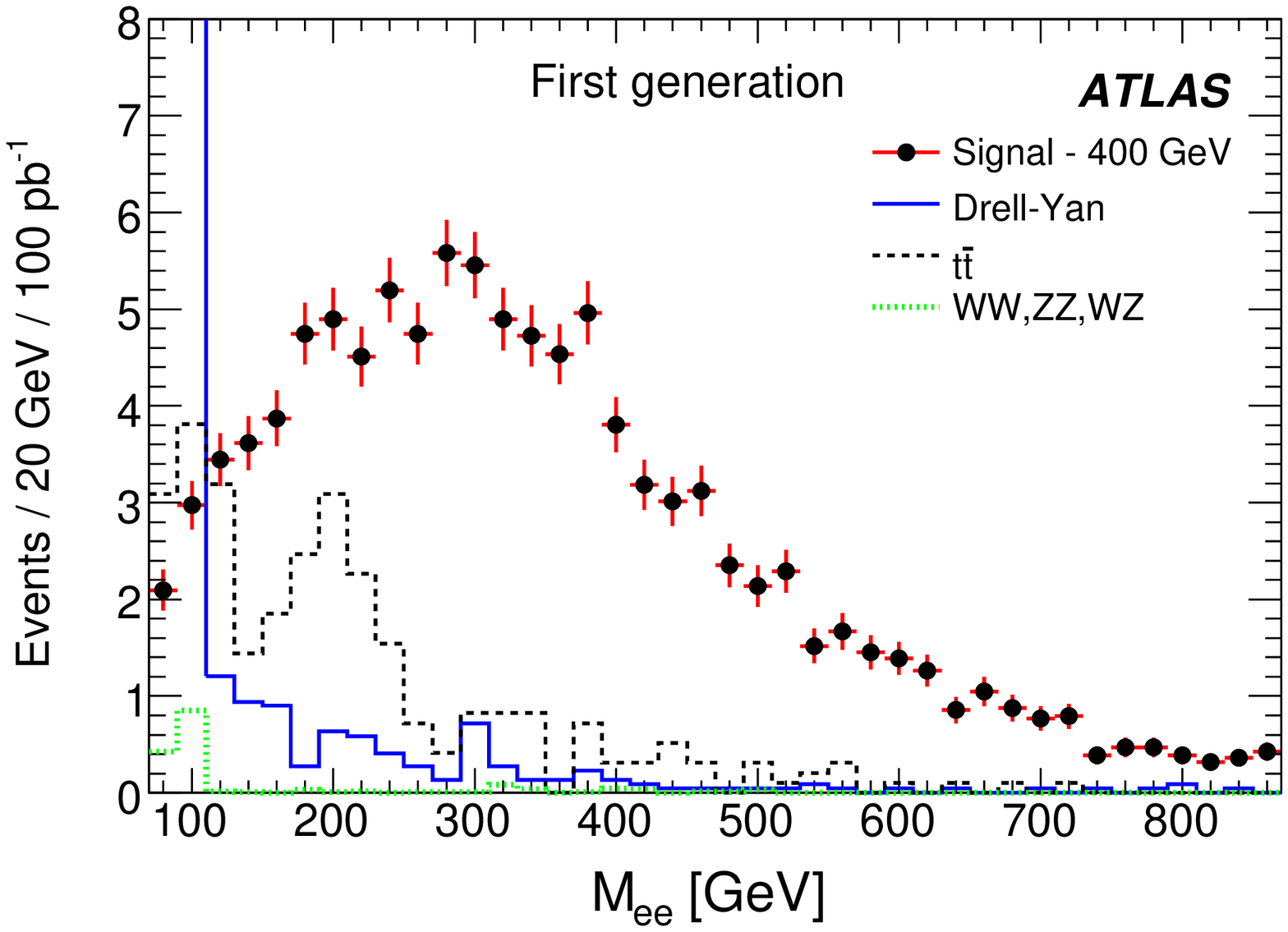}}
}
\caption{\small {
$S_{T}$ (top) and $M_{ee}$ of the selected electron pair after $S_{T}$ requirement (bottom) in 1st gen.~leptoquark MC events (m$_{LQ} = $ 400 GeV). 
Both distributions are normalized to 100~$\pbarn$ of integrated $pp$ luminosity. 
}}
\label{Lq_ee_ST_Mee_cuts}
\end{figure}

The backgrounds are suppressed and the signal significance is improved by taking advantage of the fact that the final state particles in signal-like events have relatively large $\pt$. A scalar sum of transverse momenta of signal jets and lepton candidates, denoted by $\st$, helps in reducing the backgrounds while retaining most of the signal.  The other variable used to increase the signal significance is the invariant mass of the two leptons, $M_{ll}$. The distributions of these two variables for the first generation channel are shown in Fig.~\ref{Lq_ee_ST_Mee_cuts}.

\begin{figure}[h]
\center{
{\includegraphics[width=2.9in]{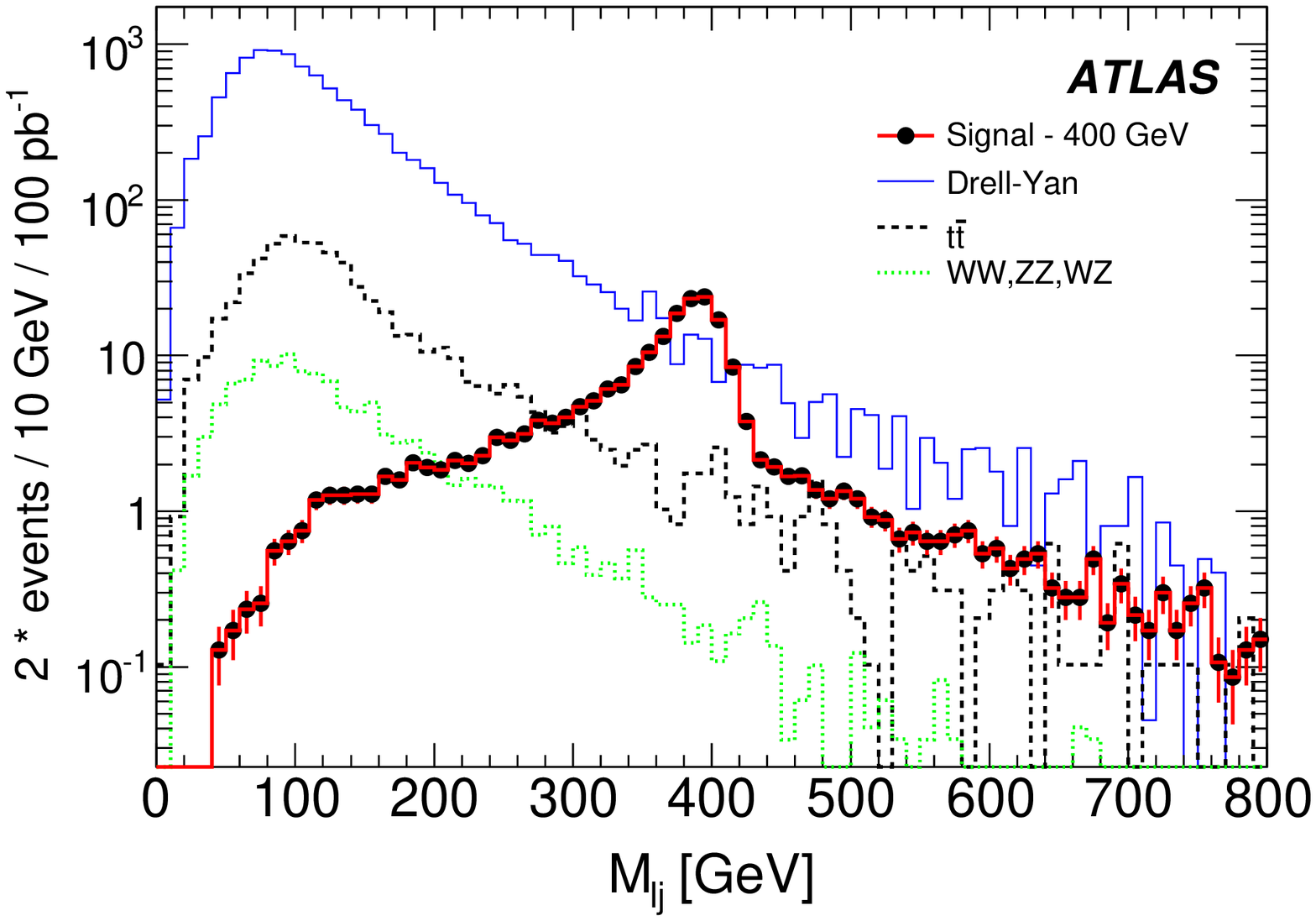}}
\hspace{0.5in}
{\includegraphics[width=2.9in]{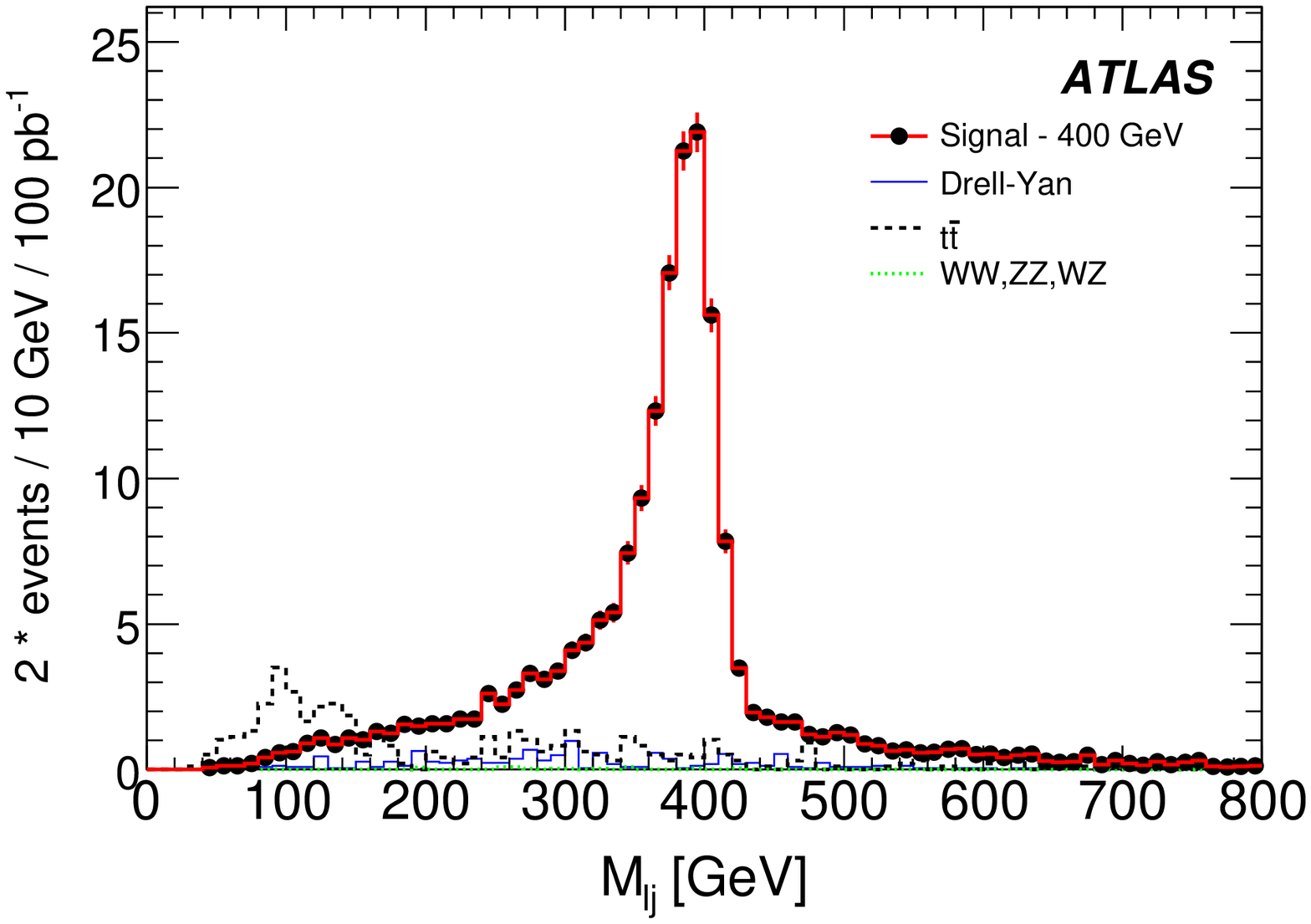}}
}
\caption{\small {
Reconstructed electron-jet invariant mass 
for 1st gen.~leptoquark (m$_{LQ}$=400~GeV) in signal 
and background MC events after baseline selection (top) and 
after additional selection criteria based on S$_T$ and m$_{ll}$ (bottom) have been applied. 
Both distributions are normalized to 100~pb$^{-1}$ of integrated $pp$ luminosity. 
}}
\label{Lq_ee_400_2D_proj_before_and_after_cuts}
\end{figure}

After applying optimized selection on these two variables, $\st$ and $M_{ll}$, relative contributions from the background processes from $t\bar{t}$, $Z /DY$, diboson and multijet are 22\%, 7\%, 0.4\% and 18\%, respectively. 
Partial cross-section for the signal and the background processes passing the selection criteria are shown in tables \ref{cut_flow_LQ_ee} and \ref{cut_flow_LQ_mm} for the first and second generation channels, respectively.
Figure \ref{Lq_ee_400_2D_proj_before_and_after_cuts} shows the invariant masses\footnote{These distributions contain two entries per event corresponding to the two reconstructed leptoquark candidates.} of the reconstructed leptoquark candidates before and after background suppression criteria are applied to the MC data.

\begin{table*}[t]
\begin{center}
\caption{
LRSM dielectron analysis. 
Partial cross-section ({\rm pb}) that remains after each selection criterion for the dielectron channel. 
} 
\begin{tabular}{|l|c|c|c|c|c|c|}
\hline
Physics              & Before     & Baseline  & $M(ejj)$        & $M(eejj)$        & $M(ee)$         & $S_T$         \\
sample               & selection  & selection & $\ge 100$~GeV   & $\ge 1000$~GeV   & $\ge 300$~GeV   & $\ge 700$~GeV \\
\hline 		                  
LRSM\_18\_3          & 0.248      & 0.0882    & 0.0882          & 0.0861           & 0.0828          & 0.0786        \\
LRSM\_15\_5          & 0.470      & 0.220     & 0.220           & 0.215            & 0.196           & 0.184         \\
\hline		                  
$Z$/DY~$\ge$~60~GeV  & 1808.      & 49.77     & 43.36           & 0.801            & 0.0132          & 0.0064        \\
$t\bar{t}$           & 450.       & 3.23      & 3.13            & 0.215            & 0.0422          & 0.0165        \\
VB pairs             & 60.94      & 0.583     & 0.522           & 0.0160           & 0.0016          & 0.0002        \\
\hline
 Multijet            & $10^{8}$   &  20.51    &  19.67          &  0.0490          & 0.0444          & 0.0444        \\
\hline
\end{tabular}
\smallskip
\label{lrsm_ee_table_selection_criteria} 
\end{center}
\end{table*}

\begin{table*}[t] 
\begin{center}
\caption{
LRSM dimuon analysis. 
Partial cross-section ({\rm pb}) that remains after each selection criterion for the dimuon channel. 
} 
\begin{tabular}{|l|c|c|c|c|c|c|}
\hline
Physics              & Before     & Baseline   & $M(\mu jj)$    & $M(\mu\mu jj)$   & $M(\mu\mu)$     & $S_T$         \\
sample               & selection  & selection  & $\ge 100$~GeV  & $\ge 1000$~GeV   & $\ge 300$~GeV   & $\ge 700$~GeV \\
\hline
LRSM\_18\_3          & 0.248      & 0.145      & 0.145          & 0.141            & 0.136           & 0.128         \\ 
LRSM\_15\_5          & 0.470      & 0.328      & 0.328          & 0.319            & 0.295           & 0.274         \\ 
\hline
$Z$/DY~$\ge$~60~GeV  & 1808.      &  79.99     & 69.13          & 1.46             & 0.0231          & 0.0127        \\
$t\bar{t}$           & 450.       &  4.17      & 4.11           & 0.275            & 0.0527          & 0.0161        \\
VB pairs             & 60.94      &  0.824     & 0.775          & 0.0242           & 0.0044          & 0.0014        \\
\hline
Multijet             & $10^{8}$   &  0.0       &  0.0           &  0.0             & 0.0             & 0.0           \\
\hline
\end{tabular} 
\label{lrsm_mm_table_selection_criteria} 
\end{center} 
\end{table*}

\subsection{Sensitivity and Discovery Potential}

\begin{figure}[h]
\center{
{\includegraphics[width=2.9in]{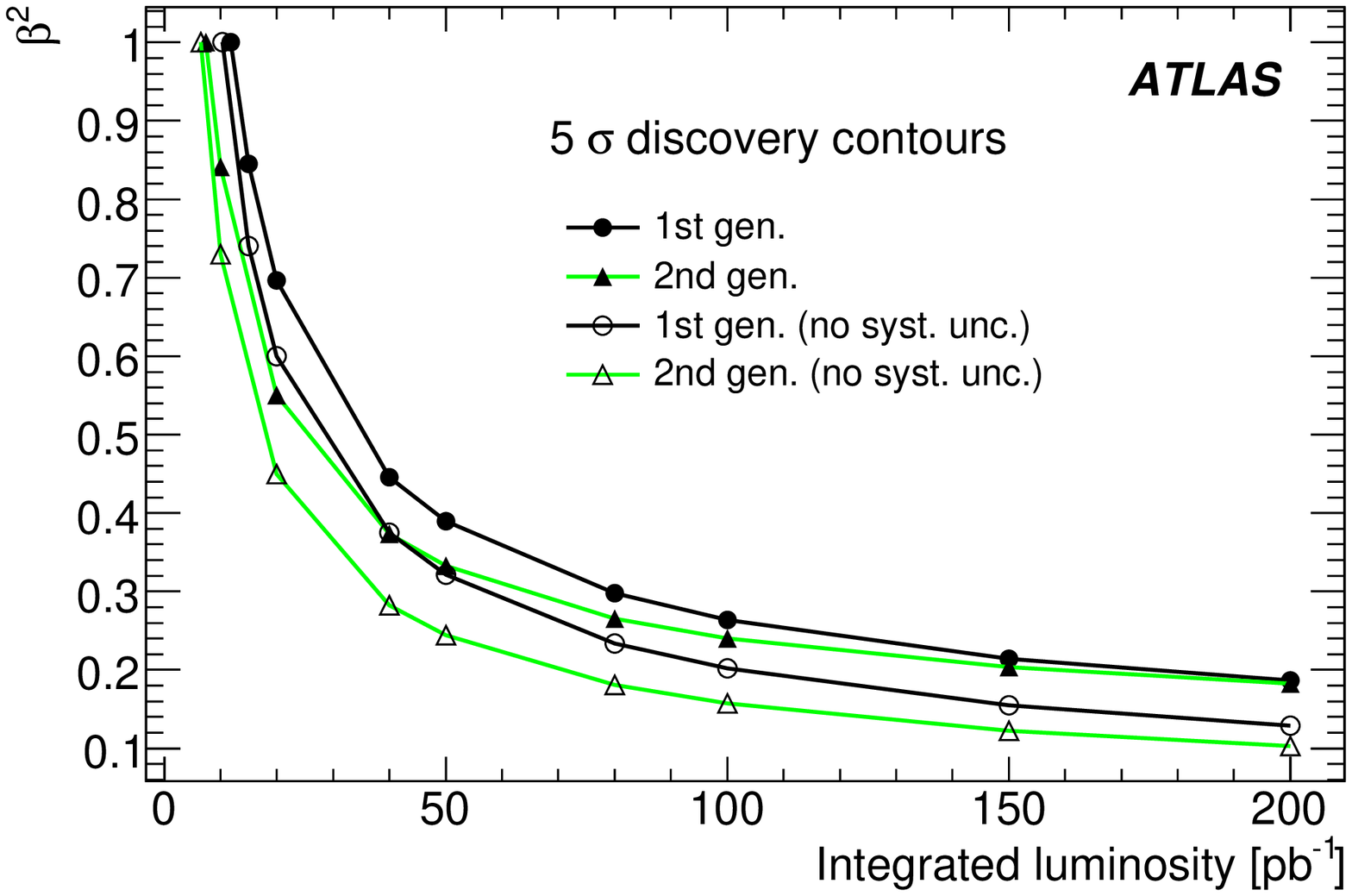}}
\hspace{0.5in}
{\includegraphics[width=2.9in]{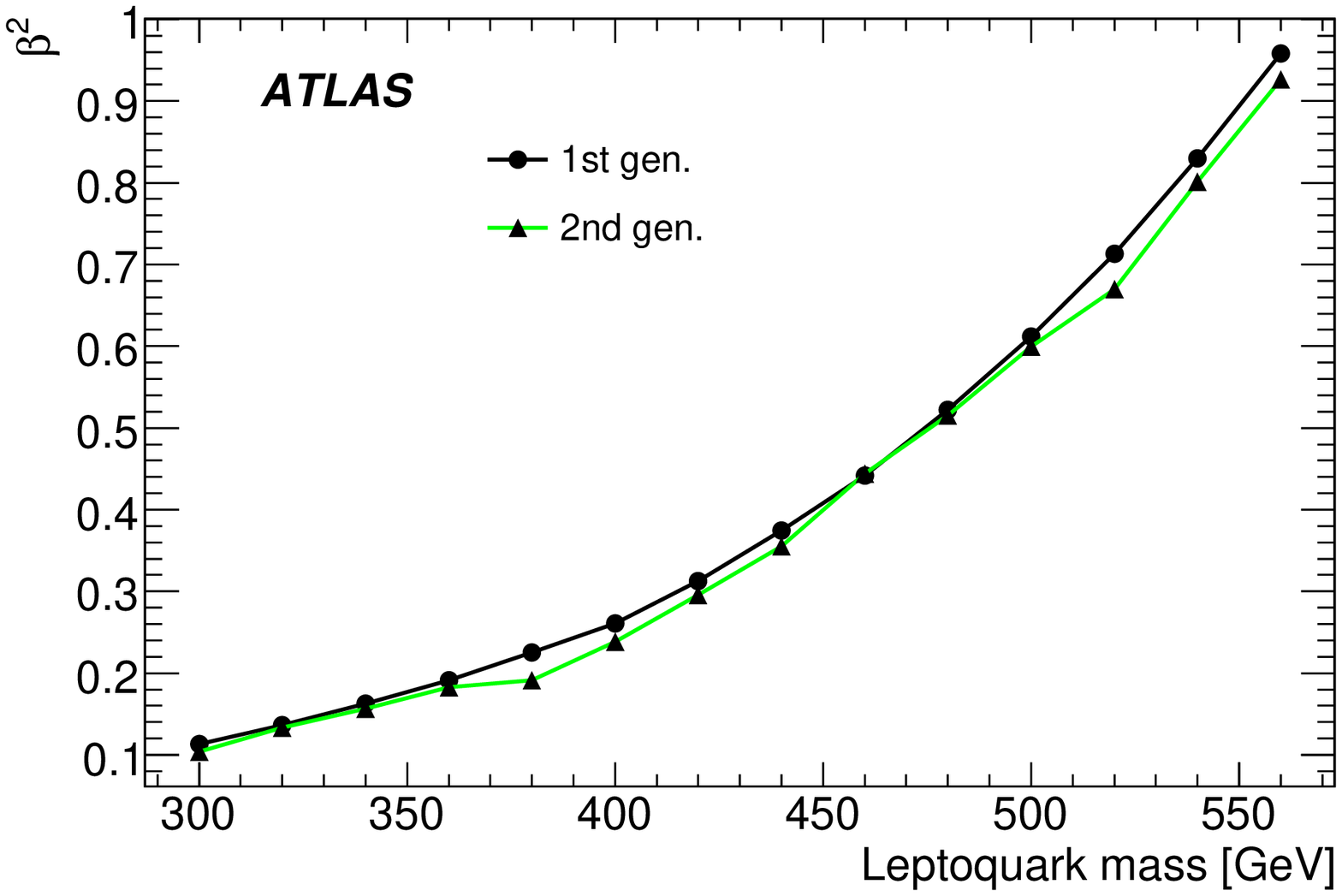}}
}
\caption{\small {
5$\sigma$ discovery potential for 1st and 2nd gen.~400~GeV  scalar leptoquark versus $\beta^2$, 
with and without background systematic uncertainty (top). Minimum $\beta^2$ of scalar leptoquark versus leptoquark mass at~100 $\pbarn$ of integrated $pp$ luminosity at 5$\sigma$ (background systematic uncertainty is included) (bottom). 
}}
\label{LQ_sensitivity}
\end{figure}

ATLAS's sensitivity to leptoquark signal for a 400 GeV mass hypothesis and with an integrated $pp$ luminosity of 100 $\pbarn$ is summarized in Fig.~\ref{LQ_sensitivity}. 
The cross-sections include systematic uncertainties of 50\%. 
Leptoquark-like events in the ATLAS detector are triggered by single leptons with an efficiency of 97\%.
ATLAS is sensitive to leptoquark masses of about 565~GeV and 575~GeV for 1st and 2nd generations, respectively, 
at the given luminosity of 100~$\pbarn$ 
provided the predicted cross-sections for the pair production of leptoquarks are correct.

\begin{figure}[h]
\center{
{\includegraphics[width=2.9in]{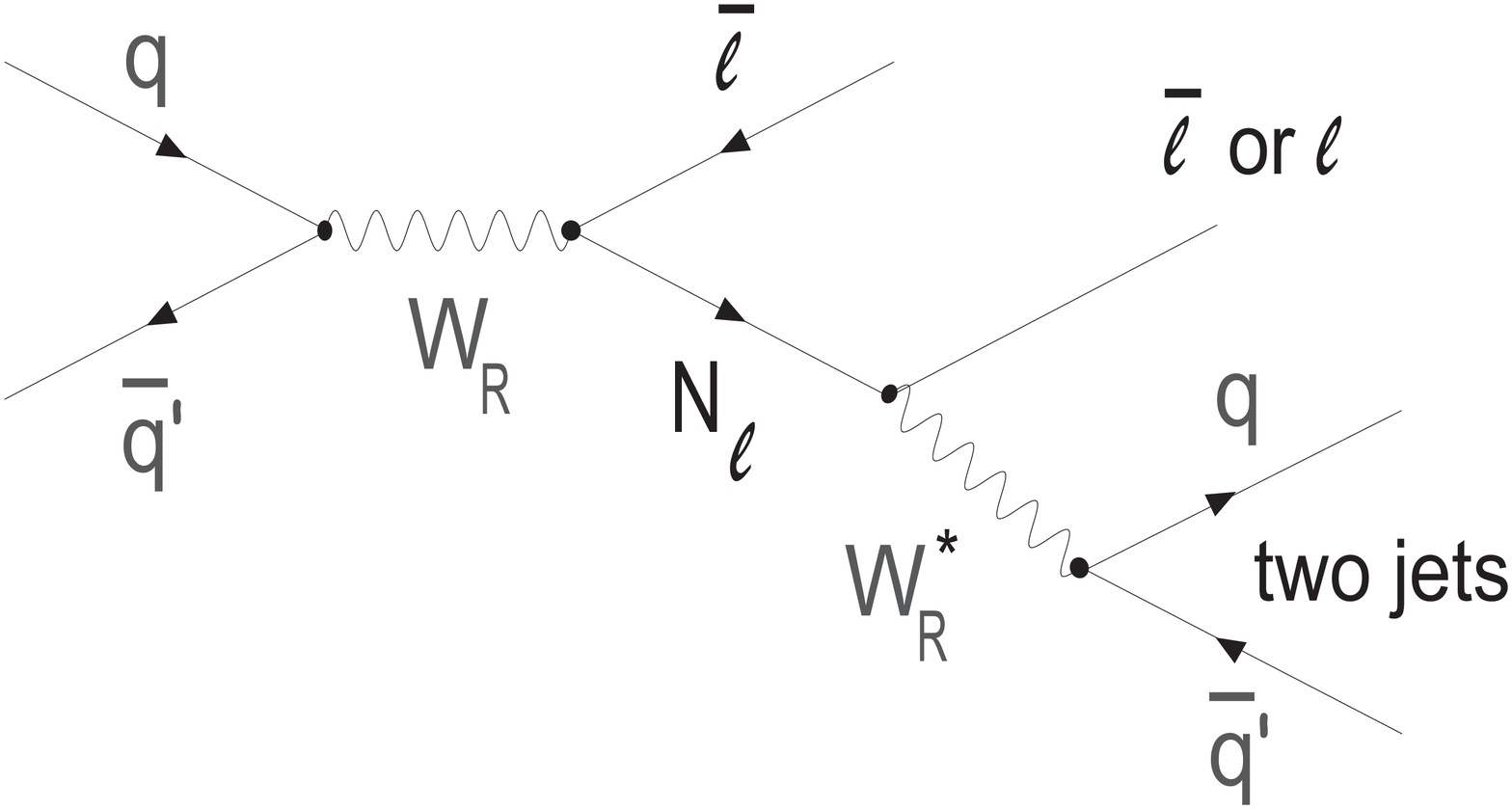}}
}
\caption{\small {
Feynman diagram for $W_R$ production and its decay to the Majorana neutrino $N_\ell$ at the LHC. 
}}
\label{lrsm_fig_feynman}
\end{figure}

\section{Search for $W_R$ bosons and heavy Majorana neutrinos}

$W_R$ bosons are the right-handed counterpart of the SM $W$ bosons. These right-handed intermediate vector bosons 
are predicted in LRSMs and can be produced at the LHC in the same processes as the SM's $W$ and $Z$. They decay into heavy Majorana neutrinos.
The Feynman diagram for $W_R$ production and subsequent decay to Majorana neutrino is shown in Fig.~\ref{lrsm_fig_feynman}.

This section describes the analysis of $W_R$ production and its decays $W_R \to e N_e$ and 
$W_R \to \mu N_\mu$, followed by the decays $N_e \to e q^\prime \bar{q}$ 
and $N_\mu \to \mu q^\prime \bar{q}$, which are detected in final states with 
(at least) two leptons and two jets. 
The two leptons can be of either same-sign or opposite-sign charge due to the Majorana nature of neutrinos. 
This analysis in both the dielectron and the dimuon channels has been performed without separating dileptons into same-sign and opposite-sign samples. 

Studies \cite{ATLAS_CSC_PHYSICS_BOOK:2008} of the discovery potential for $W_R$ and Majorana neutrinos N$_e$ and N$_{\mu}$ have been performed using
MC samples where M(N$_l$) = 300 GeV; M(W$_R$)= 1800 GeV (referred to as LRSM\_18\_3) 
and M(N$_l$) = 500 GeV; M(W$_R$) = 1500 GeV (referred to as LRSM\_15\_5), simulated with PYTHIA according to a
particular implementation \cite{A.-Ferrari:2000si} of LRSM \cite{K.-Huitu:1997ye}. 
The production cross-sections $\sigma(pp(14{\rm~TeV}) \rightarrow W_R X)$ times the branching fractions $(W_R \rightarrow l N_l \rightarrow l l j j)$ are 24.8 pb and 47 pb for  LRSM\_18\_3 and LRSM\_15\_5, respectively.

\subsection{Reconstruction and objects selection}

Signal event candidates are reconstructed using two electron or muon candidates and two
jets that pass the standard selection criteria as discussed in section \ref{baseline_selection}.
The two signal jet candidates are combined with each of the signal leptons and the combination that gives
the smaller invariant mass is assumed to be the new heavy neutrino candidate. 
The other remaining lepton is assumed to come directly from the decay of the $W_R$ boson.
If signal electrons and signal jets overlap in $\Delta R$ within 0.4 then, 
to avoid double counting, 
only the two signal jets are used to reconstruct the invariant masses of the heavy neutrino candidate and $W_R$. 

\subsection{Background Studies}

\begin{figure}[htbp]
\center{
{\includegraphics[width=2.9in]{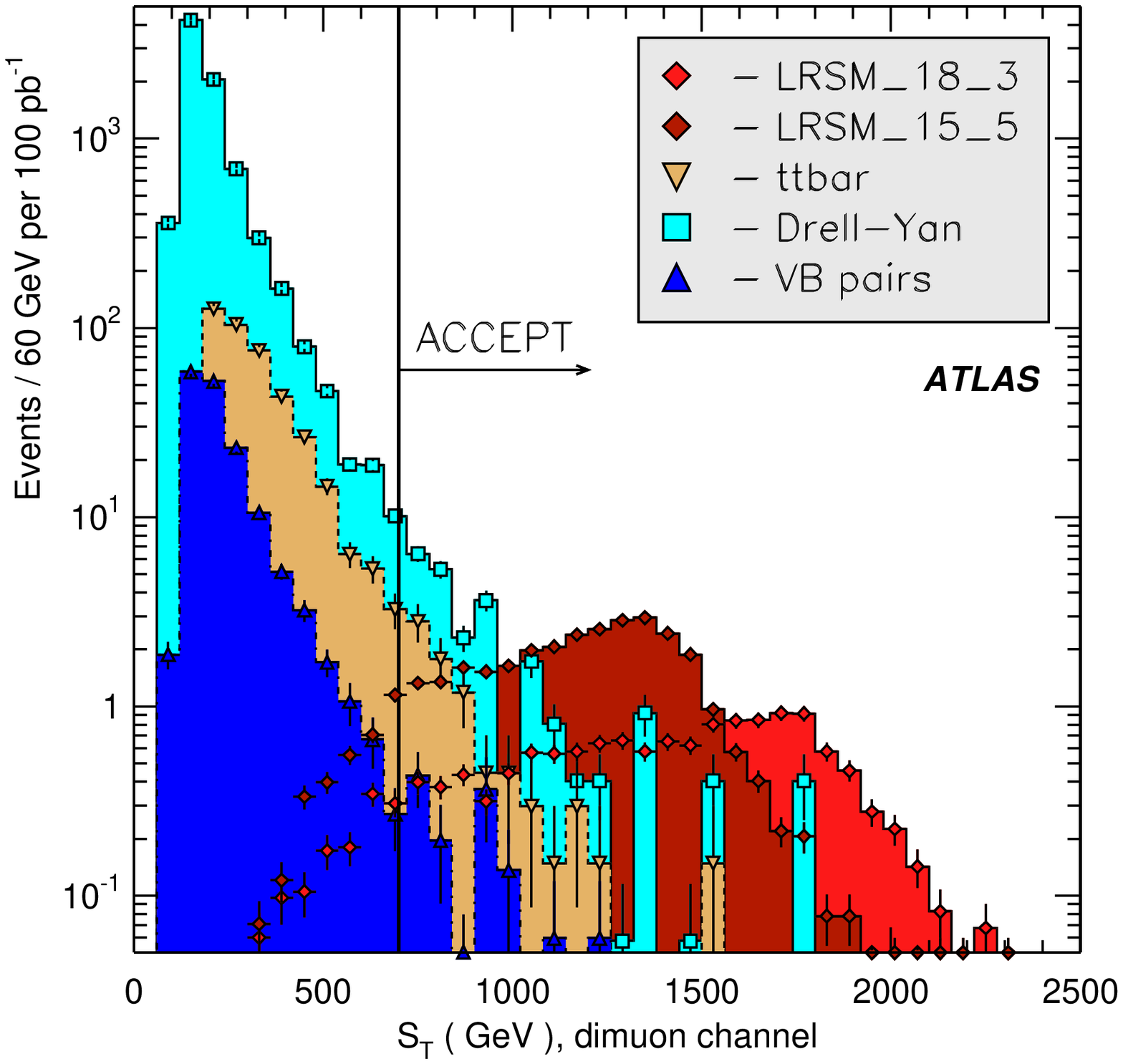}}
{\includegraphics[width=2.9in]{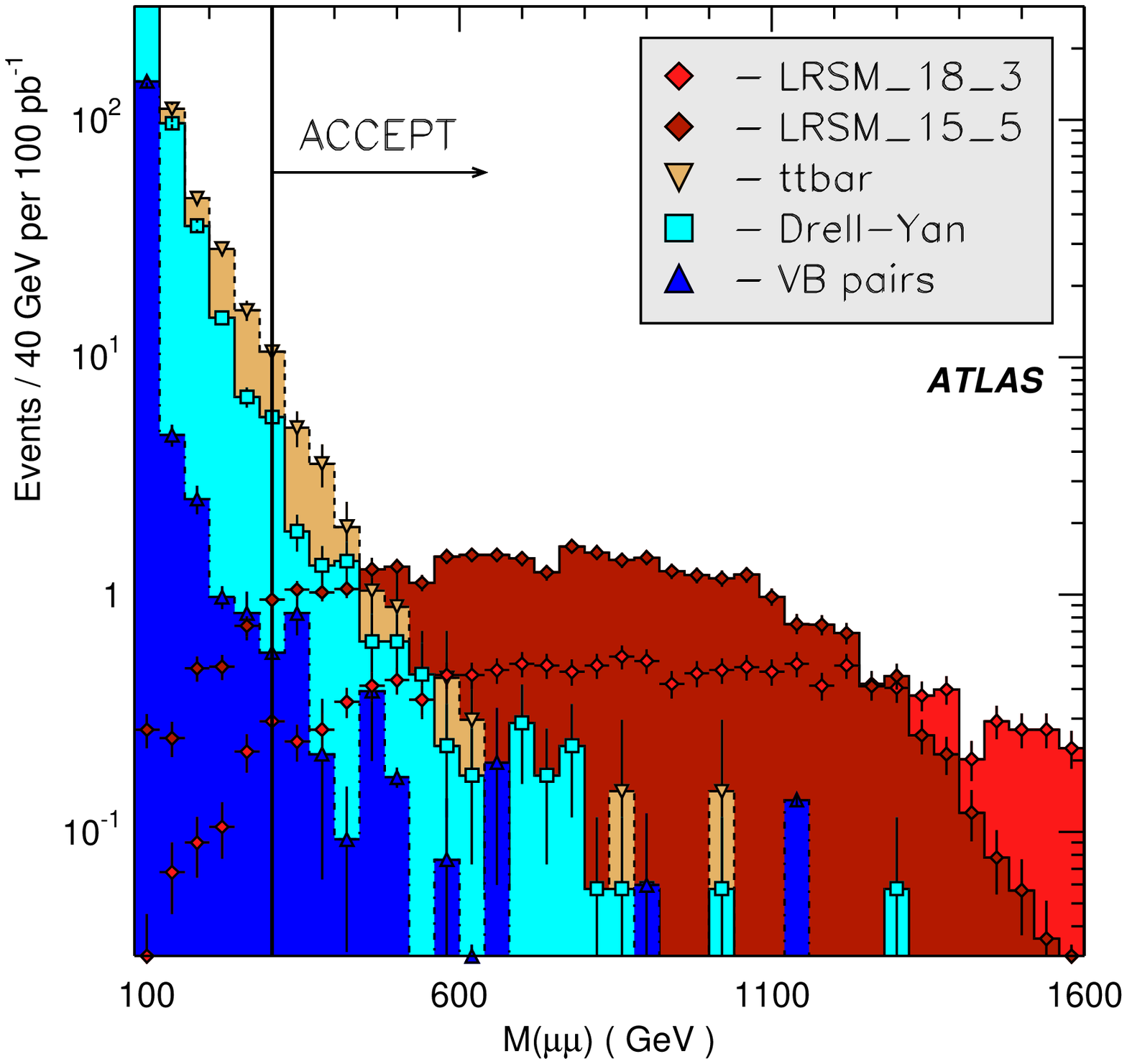}}
}
\caption{\small {
LRSM analysis. 
The distributions of $S_T$ (top) and $M(\ell \ell)$ (bottom) for signals and backgrounds 
normalized to 100${\rm pb^{-1}}$ of integrated $pp$ luminosity 
after baseline selection in the dimuon analysis. 
}}
\label{lrsm_mm_fig_st_mll}
\end{figure}

\begin{figure}[h]
\center{
{\includegraphics[width=2.9in]{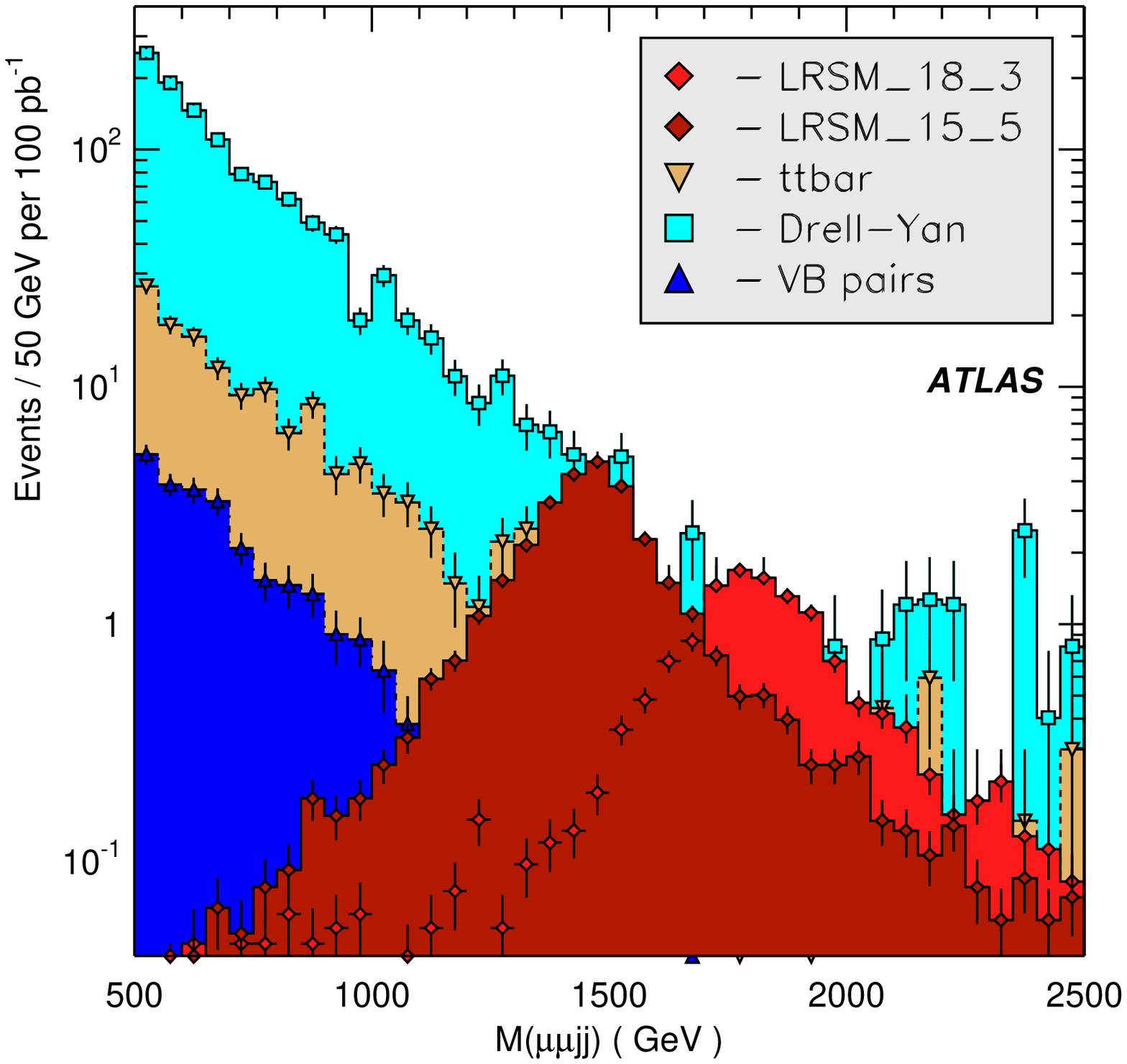}}
{\includegraphics[width=2.9in]{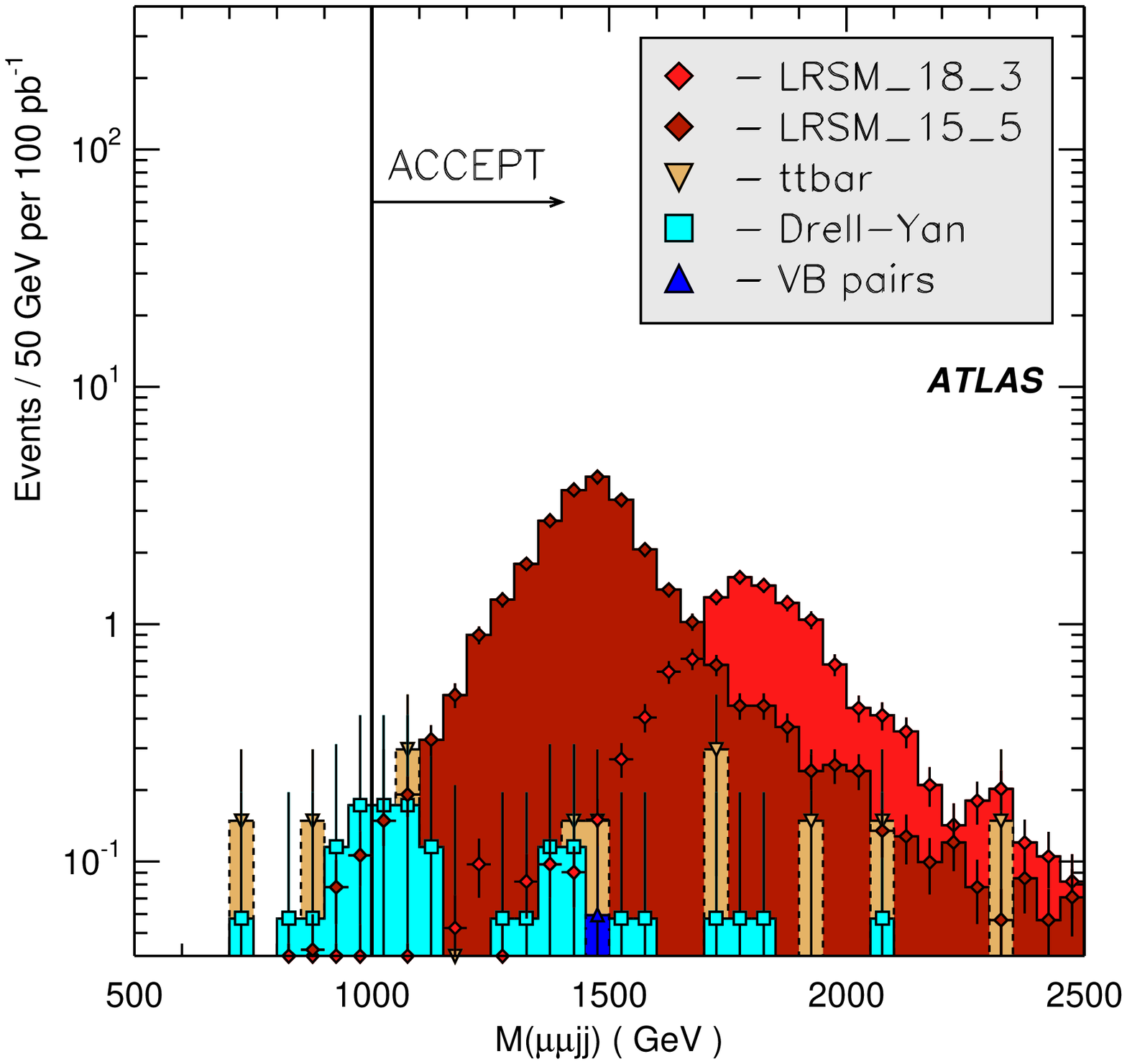}}
}
\caption{\small {
LRSM analysis. 
The distributions of the reconstructed invariant masses for 
$W_R \to \mu N_\mu$ candidates 
in background and signal (LRSM\_18\_3 and LRSM\_15\_5) events 
before (top) and after (bottom) background suppression is performed in dimuon channel analysis. 
Both distributions are normalized to $100 {~\rm pb^{-1}}$ of integrated $pp$ luminosity. 
}}
\label{lrsm_mm_fig_wr_masses}
\end{figure}

The main backgrounds to the LRSM analyses studied here are the same as mentioned in section \ref{LQ_Bckg}. 
The same background suppression criteria as in the leptoquark analyses are also effective here, namely S$_T$ and m$_{ll}$.
The distributions of these two variables for the dimuon channel are shown in Fig.~\ref{lrsm_mm_fig_st_mll}.
Partial cross-section for the signal and the background processes passing the selection criteria are shown in tables \ref{lrsm_ee_table_selection_criteria} and \ref{lrsm_mm_table_selection_criteria} for the dielectron and dimuon channels, respectively.
Figure \ref{lrsm_mm_fig_wr_masses} shows the invariant mass of the reconstructed $W_R$ candidates before and after background suppression criteria are applied to the MC data.

\subsection{Sensitivity and Discovery Potential}

\begin{figure}[b]
\center{
{\includegraphics[width=2.9in]{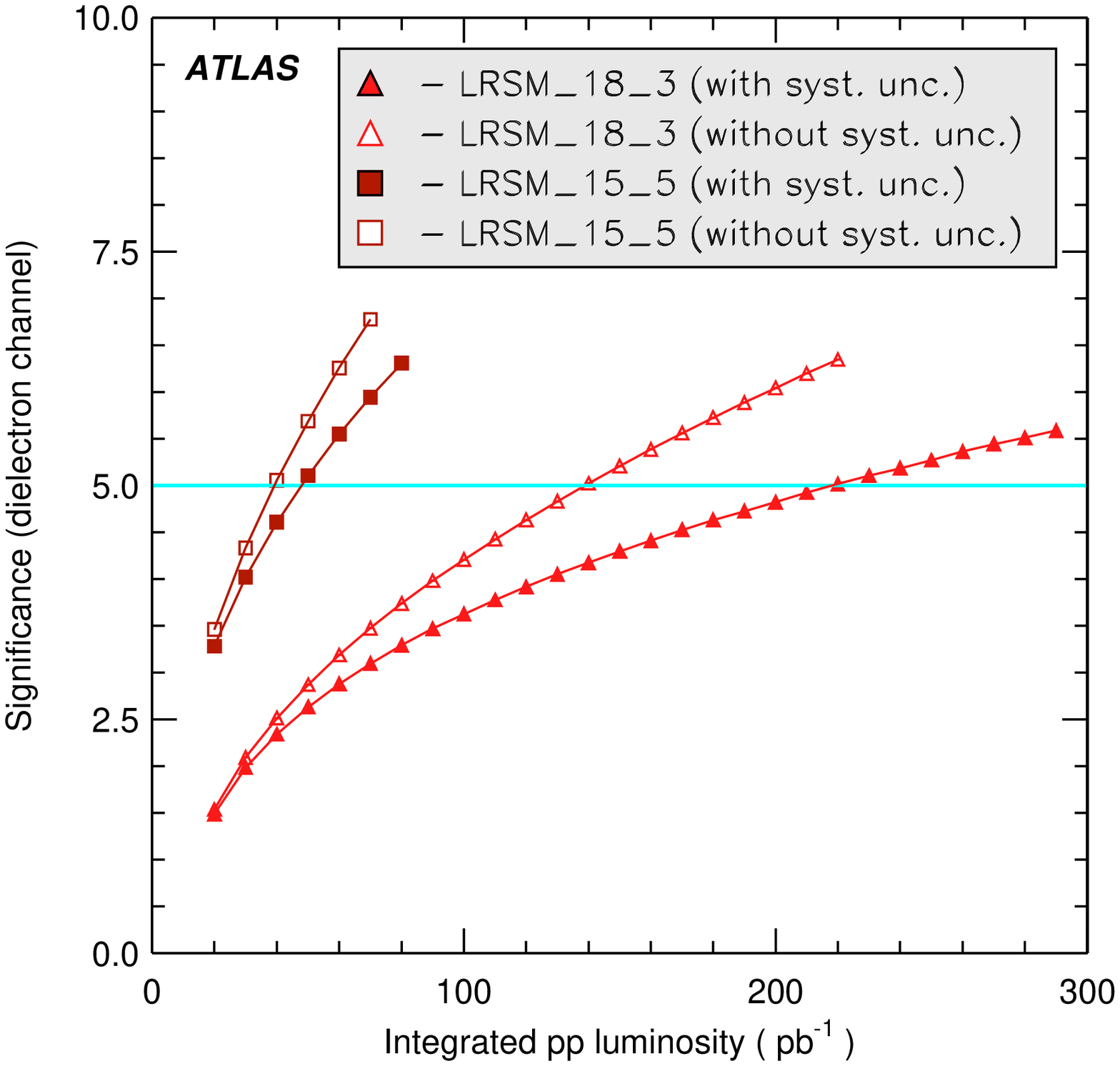}}
{\includegraphics[width=2.9in]{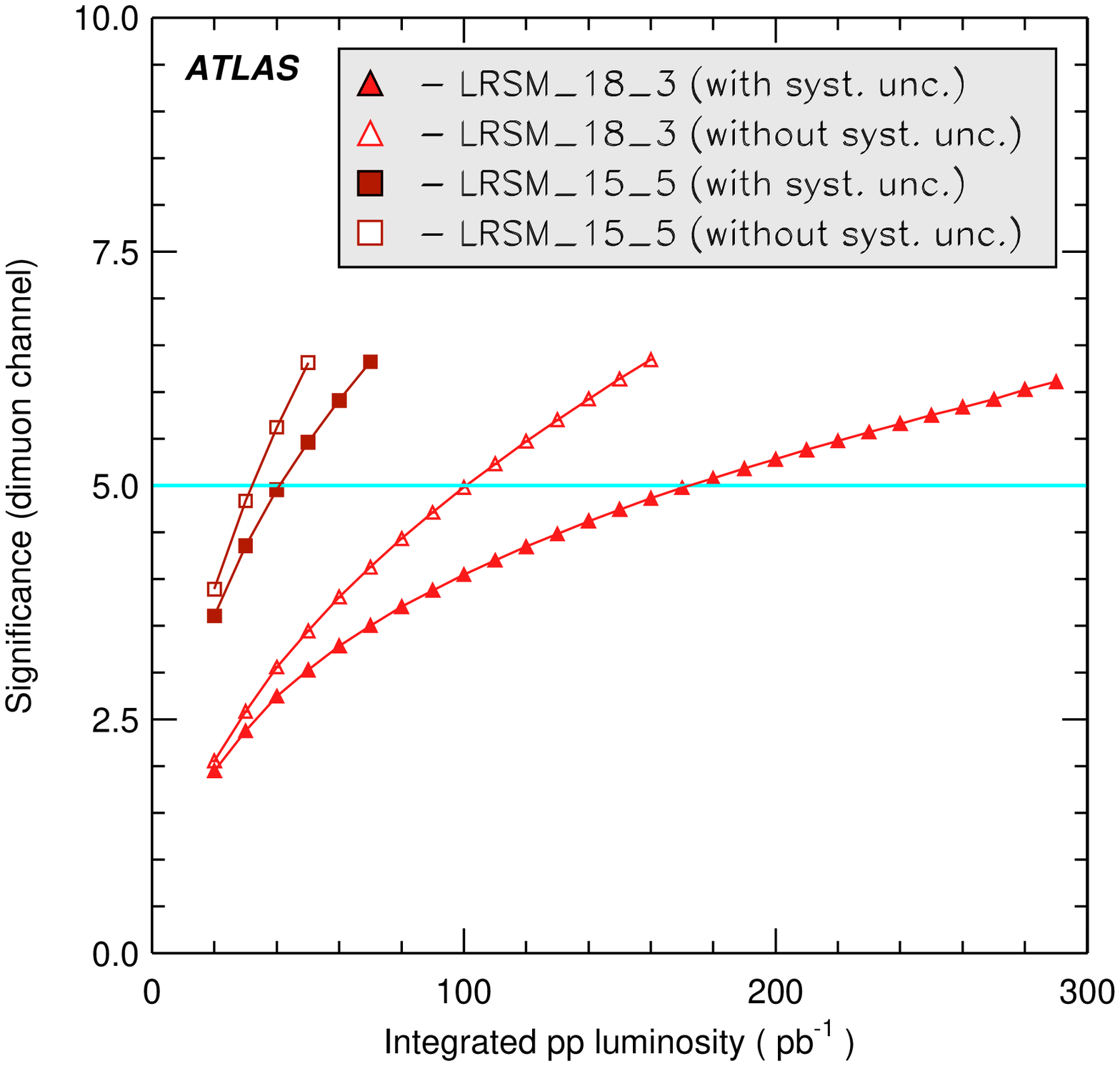}}
}
\caption{\small {
LRSM analysis. 
Expected signal significances versus integrated $pp$ 
luminosity for $N_e$, $N_\mu$ and $W_R$ mass hypotheses, 
according to signal MC samples LRSM\_18\_3 and LRSM\_15\_5. 
Open symbols show sensitivities without systematic uncertainties. 
Sensitivities shown with closed symbols include an overall relative uncertainty 
of 45\% (40\%), estimated for background contributions in the dielectron (dimuon) analysis. 
}}
\label{lrsm_fig_discovery}
\end{figure}

Signal significance for $W_R$ analyses in the dielectron and dimuon channels as a function of integrated $pp$ luminosity at 14~TeV is summarized in Fig.~\ref{lrsm_fig_discovery}. The results include systematic uncertainty of 45\% and 40\% for dielectron and dimuon channel, respectively. The events in this analysis are also triggered by single leptons with an efficiency of 97\%. 

\section{Conclusions}

Dilepton-jet based final states have been discussed in both electron and muon channels.
Discovery potential for leptoquarks and LRSM with early LHC data have been investigated 
with the predicted cross-sections for these models.
Assuming a $\beta$ = 1, both 1st  and 2nd generations leptoquarks could be discovered 
with masses up to 550~GeV with 100~${\rm pb^{-1}}$ of data.
Two LRSM mass points LRSM\_18\_3 and LRSM\_15\_5
for the $W_R$ bosons and heavy Majorana neutrinos have been
studied. The discovery of these new particles with such 
masses would require integrated luminosities of 
150~${\rm pb^{-1}}$and 40~${\rm pb^{-1}}$, respectively.

\bigskip 

\end{document}